\documentclass[a4paper]{spie}  %>>> use this instead for A4 paper
\usepackage{amsmath,amsfonts,amssymb}
\usepackage[colorlinks=true, allcolors=blue]{hyperref}

\usepackage[english]{babel}
\usepackage[utf8]{inputenc}
\usepackage[]{graphicx}
\usepackage{microtype}
\usepackage{gensymb}
\usepackage{epsfig}
\usepackage{float}
\usepackage{subfig}
\usepackage{gensymb}
\usepackage{todonotes}
\usepackage{boxedminipage2e}
\usepackage{siunitx}
\usepackage{booktabs}
\usepackage{wasysym}

\usepackage{natbib}
\bibpunct{(}{)}{;}{a}{}{,}

\graphicspath{{Pics/}}
\usepackage{txfonts}
\usepackage{enumerate}

\usepackage{pgfplots}
\pgfplotsset{compat=newest}
\usepgfplotslibrary{fillbetween}
\usetikzlibrary{backgrounds}

\title{Observing the integrated and spatially resolved Sun with ultra-high spectral resolution }

\author[a]{Sch\"afer, S.}
\author[a]{Royen, K.}
\author[a]{Huster Zapke, A.}
\author[a]{Ellwarth, M.}
\author[a]{Reiners, A.}

\affil[a]{Georg-August Universit\"at G\"ottingen, Institut f\"ur Astrophysik, Friedrich-Hund-Platz 1, 37077 G\"ottingen, Germany}
\authorinfo{Further author information: (Send correspondence to Sebastian Sch\"afer)\\ Sebastian Sch\"afer: E-mail: schaefer@astro.physik.uni-goettingen.de, Telephone: +49 (0)551 39 25068}

\begin{document} 
\maketitle

\begin{abstract}

The Institute for Astrophysics Göttingen operates a solar observatory that combines a 50\,cm siderostat with (1) a vacuum vertical telescope, (2) a very high resolution Fourier Transform Spectrograph ($R > 900,000$ at 600\,nm), and (3) a Laser Frequency Comb for extremely precise and accurate frequency calibration ($<10\,cm/s$). We introduce our setup that feeds the spectrograph with either a 32.5" field of view of the solar surface, or with disk-integrated sunlight for Sun-as-a-star observations and explain the necessary computational steps to guide specific positions on the solar surface into the fiber.

Our instrument suite can deliver spectroscopic measurements with extremely accurate frequency calibration, which is valid across very large frequency regions (approx. 400-800\,nm in wavelength). This allows precision spectroscopy of individual lines in order to study the variability of spectral lines in Sun-as-a-star observations as well as determining the convective blueshift across the solar surface from many spectral lines.
\end{abstract}

\keywords{Sun, Spectroscopy, Fiber, Radial-Velocity, Guiding}

\section{Motivation}

The Sun provides invaluable reference for understanding the
influence of stellar surface velocity fields and active regions
because it is possible to relate precision radial velocity (RV) observations to
spatially resolved solar surface information
\citep{2010A&A...519A..66M, 2016MNRAS.457.3637H}. However, obtaining solar disk observations with m\,s$^{-1}$
precision or better is
extremely difficult, and current solar Dopplergrams are known to
exhibit severe problems in this respect \citep{2012SoPh..275..285C,
  2013ApJ...765...98W, 2013A&A...551A.105L}. Furthermore, measuring
precise Sun-as-a-star RVs is very challenging because the spatial
extension of the Sun implies that feeding light from the Sun into a
spectrograph involves the problem of collecting light from all areas
of the solar disk equally. Earlier attempts have demonstrated this
difficulty \citep{Jimenez1986AdSpR, Deming1987ApJ, Deming1994ApJ,
  McMillan1993ApJ} but there is new motivation for instruments facing
this challenge, for example, observing the Sun with HARPS
\citep{2015ApJ...814L..21D} and in the G\"ottingen Solar Radial
Velocity Project \citep{2016PASP..128i5002L}. For example, disk-integrated data taken with HARPS have demonstrated its high potential for understanding stellar RV variability \citep{2019MNRAS.487.1082C} and also for understanding systematic instrumental effects \citep{2020arXiv200901945D}.

The main mechanism causing RV variability in stellar observations is the influence of magnetic activity on convective blueshift
\citep{2010A&A...512A..38L, 2010A&A...512A..39M, 2010A&A...519A..66M,
  2010A&A...520A..53L, 2014ApJ...796..132D, 2015ApJ...798...63M,
  2016MNRAS.457.3637H}. The motion of hot, rising plasma on the
surface of a star causes a Doppler shift of the observed spectrum
\citep{1978SoPh...58..243B, 1981A&A....96..345D,
  1982Natur.297..208L}. Because the motion of the plasma is
non-isotropic, the amplitude of the Doppler shift varies across the
stellar disk (it depends on the stellar limb angle). For a Sun-like
star, the disk-integrated effect is on the order of a hundred
m\,s$^{-1}$ such that the observed radial velocity of a star is
systematically blueshifted. The variability of RV observations is then
caused by the presence of temporally evolving and co-rotating active
regions. In an active region, the magnetic field can inhibit
convection and shut off convective blueshift with the result that the
spectrum emerging from that area appears redshifted with respect to
the quiet photosphere.

Our motivation for observing the Sun with a Fourier Transform Spectrograph (FTS) is twofold. First, an FTS provides very high spectral resolution (up to $10^6$) allowing to study effects that cannot be resolved in typical astronomical spectrographs. Our dataset is complementary to data from other facilities because we are targeting the analysis of spectral lines at the highest resolution across very large wavelength ranges but at lower cadence than other programs. Second, frequency calibration of an FTS in principle is extremely robust because light of all frequencies follows the same optical path, which allows to use light with a limited frequency range for calibrating the entire spectrum. While this does not render frequency calibration an easy task, it eliminates a large number of free parameters inherent in \'echelle spectrograph calibration.

\section{Overview of instrumental setup at IAG}
\begin{figure}[H]
  \centering
  \fbox{\includegraphics[width=1\linewidth]{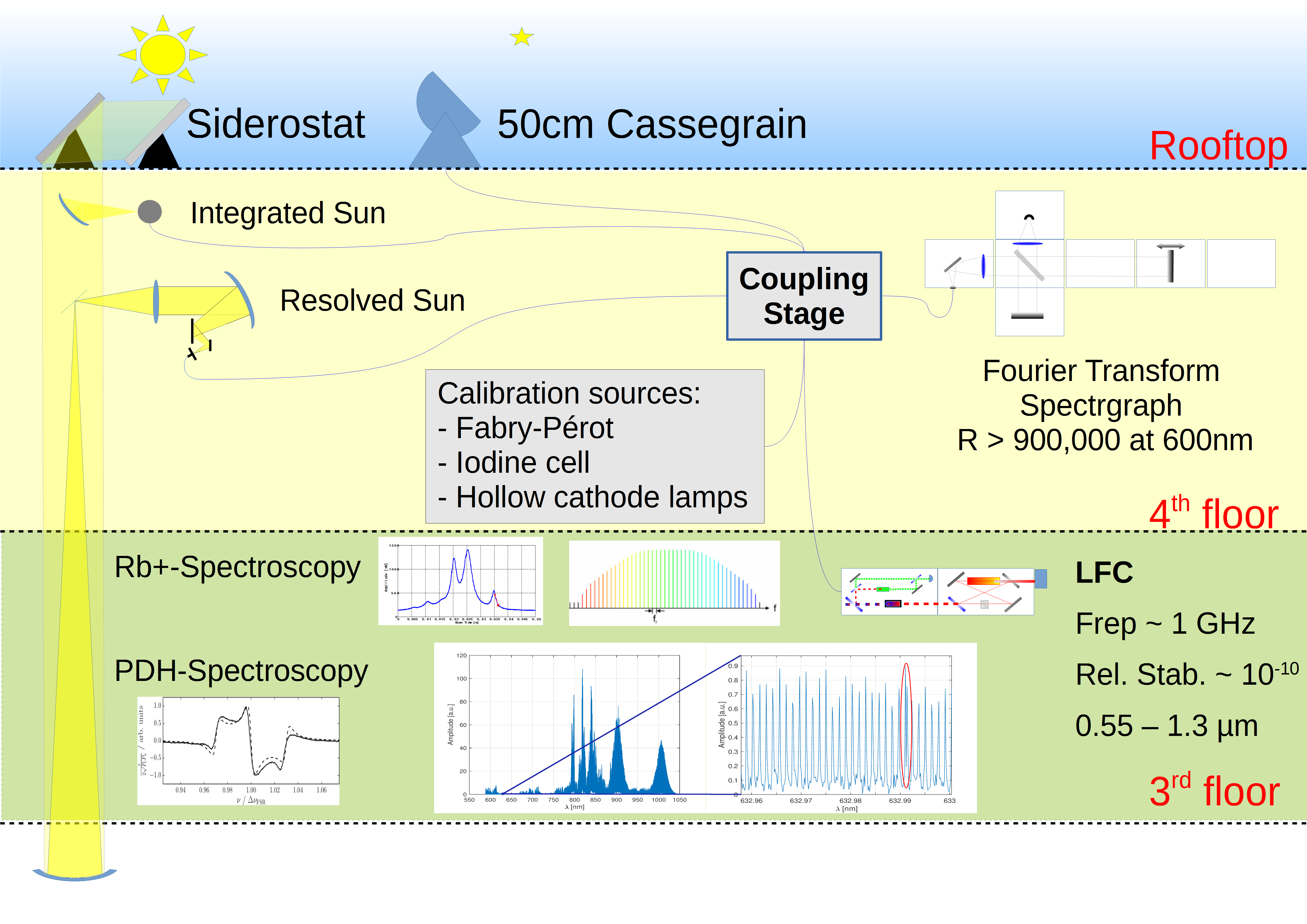}}
  \caption{Rooftop: 50\,cm siderostat with vertical vacuum telescope, integrated Sun setup with integrating sphere, 50\,cm Cassegrain telescope for stellar observations; 4th floor: Resolved Sun setup, Fourier Transform Spectrograph (fiber-fed, $R>900,000$), calibration sources; 3rd floor: Laser Frequency Comb.}
 \label{fig:IAG}%
\end{figure}

The Institute for Astrophysics Göttingen (IAG) is hosting a number of instruments: On the rooftop there are the 50\,cm Cassegrain night telescope and a 50\,cm siderostat for solar observations. Attached to the siderostat is an integrated Sun setup to observe the Sun as a star (see Sec.\,\ref{Sec:integrated}). The siderostat is connected to a vacuum tube, reaching down to below the 3rd floor where the primary mirror is located, making it a vertical vacuum telescope (see Sec.\,\ref{Sec:Telescope} for more details). The focal point of the Solar telescope is inside the optical lab on the 4th floor, where the resolved Sun setup is located (see Sec.\,\ref{Sec:resolved}). This lab also hosts a Fourier Transform Spectrograph (Sec.\,\ref{Sec:FTS}), a number of calibration sources and a coupling stage with two fibers leading into the FTS. Finally in the 3rd floor lab a Laser Frequency Comb is located, with fibers connecting it to the 4th floor lab. Figure\,\ref{fig:IAG} shows a crosssection of the building (not to scale) to illustrate the location of the different setups and their connections.

\section{Telescope}
\label{Sec:Telescope}
\begin{figure}[H]
  \centering
  \fbox{\includegraphics[width=0.6\linewidth]{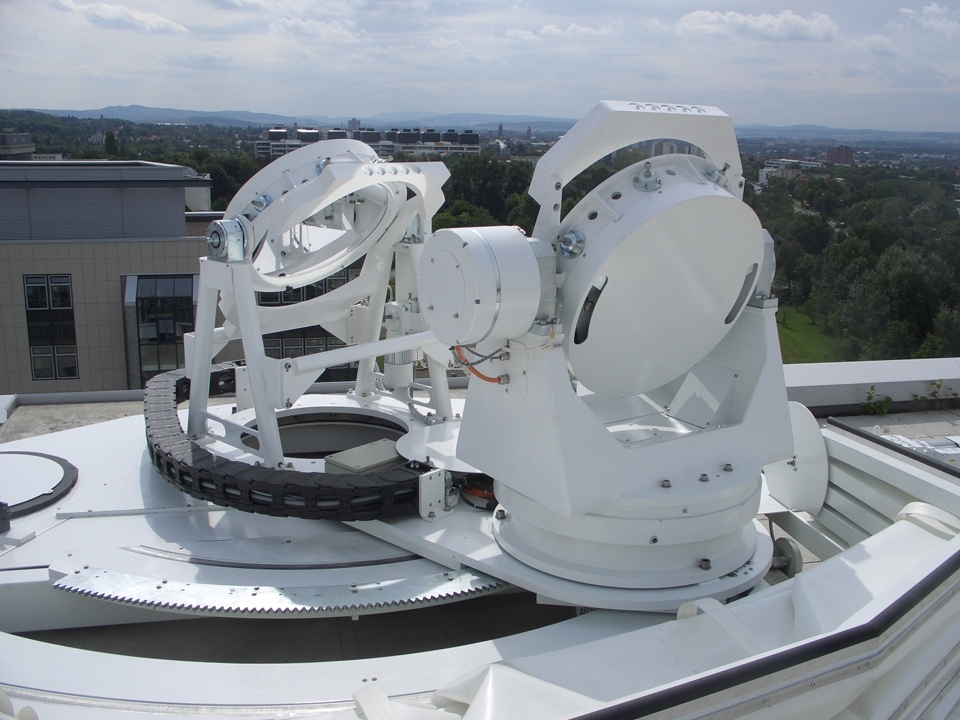}}
  \caption{Siderostat on the roof, view southbound towards the city center of Göttingen: The S1 mirror on the right is tracking the movement of the target object (e.g. the Sun), the S2 mirror on the left is fixed in its 45\degree position. The  vacuum window of the vertical part of the telescope is  below the S2.  }
 \label{fig:Roof}%
\end{figure}

The siderostat has been built by Halfmann Teleskoptechnik. A vertical vacuum tube was built  directly into the new phycics building with the siderostat on top of the 5th floor and the primary mirror located below the 3rd floor. Installation of the system and first light happened in September 2008 and for a couple of years the solar telescope was mainly used for student lab courses, featuring an open air Littrow spektrograph with a focal length of 8\,m and a resolving power of over 100,000 in a small wavelength range. This allowed the observation of sunspots and the measurement of their magnetic field strength across the slit of the spectrograph via Zeeman splitting. Figure\,\ref{fig:Roof} shows the siderostat with the view towards Göttingen and Table\,\ref{tab:mirrors} gives an overview over the mirrors used.

\begin{table}[H]
\caption{Telescope mirror system} 
\label{tab:mirrors}
\begin{center}       
\begin{tabular}{c|c|c|c|c} 
\hline
\rule[-1ex]{0pt}{3.5ex}  & S1 & S2 & M1 & Nasmyth   \\
\hline
\rule[-1ex]{0pt}{3.5ex}  Type & flat circular  & flat circular  & on axis parabolic  & elliptical 45\degree flat    \\
\hline
\rule[-1ex]{0pt}{3.5ex}   Diameter & {\begin{tabular}{c} 515.3\,mm (mech)\\ 502\,mm (opt) \end{tabular}} & {\begin{tabular}{c} 515.3\,mm (mech)\\ 502\,mm (opt) \end{tabular}} & 500\,mm & 120\,mm (semi-minor axis)   \\
\hline
\rule[-1ex]{0pt}{3.5ex}  Thickness & 63.2\,mm & 62.8\,mm & & 25\,mm  \\
\hline
\rule[-1ex]{0pt}{3.5ex}  Focal length & - & - & 5750\,mm &   \\
\hline
\rule[-1ex]{0pt}{3.5ex}  P-V wavefront error & 0.118$\lambda$ & 0.115$\lambda$ & 0.1$\lambda$ & 0.1$\lambda$  \\
\hline
\rule[-1ex]{0pt}{3.5ex}  RMS wavefront error & 0.024$\lambda$ & 0.021$\lambda$& 0.1$\lambda$ &  0.1$\lambda$ \\
\hline
\rule[-1ex]{0pt}{3.5ex}  {\begin{tabular}{c} Surface quality\\ (scratch/dig) \end{tabular}} & 70/50 & 80/50 & &   \\
\hline
\rule[-1ex]{0pt}{3.5ex}  Material & Zerodur &  Zerodu & Zerodur & Zerodur  \\
\hline
\rule[-1ex]{0pt}{3.5ex}  Coating & Al+SiO2 & Al+SiO2 & &   \\
\hline
\rule[-1ex]{0pt}{3.5ex}  Reflectivity & 90.2\% & 90.3\%& &   \\
\end{tabular}
\end{center}
\end{table}

\section{Spectrograph}

\label{Sec:FTS}
Our spectrograph is a Bruker IFS125HR Fourier-Transform-Spectrograph (FTS). In principle the FTS is a Michelson-Interferometer: The light is divided by a beamsplitter, one part being back-reflected by a fixed mirror and the other part by a movable mirror. The beamsplitter recombines the light and the resulting interference signal is captured by a detector as a function of the position of the movable mirror. This is called the interferogram. Performing a fast Fourier transform on the interferogram results in the spectrum of the input light. This can either be done with the OPUS software package (that is also controlling the FTS) or manually.

For our solar observations we use the double-sided mode, which means we can achieve a resolving power of up to $R=925,000$ at 600\,nm. In contrast to Echelle spectrographs the FTS doesn't have an exposure-time. Instead the movable mirror has to scan the whole optical path difference of the Michelson-Interferometer to achieve a certain resolution. A single scan at maximum resolution takes about 1.3\,minutes. The equivalent of taking a longer exposure  time is to do multiple scans and adding them up. This increases the signal to noise ratio similarly to a longer exposure time on an Echelle spectrograph. 

We modified our FTS so that we can now input two light sources simultaneously - a 'science light' and a calibration source, a technique commonly used in high precision spectroscopy, e.g. HARPS \citep{2003Msngr.114...20M}, CARMENES \citep{2012SPIE.8446E..0RQ}. Even though the FTS has an internal HeNe laser, stabilized to its TEM01 mode, external drift calibration is needed to achieve a radial velocity stability of better than $\sim$ 10\,m/s. Available options for our FTS include Hollow cathode lamps, a Fabry-Pérot Etalon or a Laser Frequency comb. 

For more details about our FTS modifications and the dual-source setup see \cite{Schaefer2020a}. A detailed characterization of the FTS using our laser frequency comb can be found in \cite{Huke:19}

\hspace{15cm}

\section{Resolved Sun Setup}
\label{Sec:resolved}

Initially the Solar telescope was used with a Gregory setup: The primary mirror (M1) produces an image of the Sun onto the 45\degree  Nasmyth mirror which has a conic hole in the direction of the optical axis. Above the Nasmyth the Gregory mirror is placed. It is re-imaging the Sun onto a spectrograph slit via the Nasmyth's backside. This setup is no longer used, but the hole in the center of the Nasmyth is still there and is marking the center of the telescope axis (see Fig.\,\ref{fig:Sun1}). 

In order to use the FTS to do spectroscopic observations of the spatially resolved Sun we build a new setup: The primary image of the Sun on the front side of the Nasmyth is used and re-imaged onto a fiber pickup. The fiber is leading towards the FTS. In order to select different observation regions on the Sun the siderostat mirror S1 is moved until the region of interest is placed onto the fiber. 

Section\,\ref{Sec:resolved_optic}\, explains the optical setup and  Sec.\,\ref{Sec:resolved_results} shows first results. The computational steps needed to move the telescope correctly in order to track the region of interest inside the fiber follow in the next section (Sec.\,\ref{Sec:Software}). 

\subsection{Optical setup}
\label{Sec:resolved_optic}

The current version of the resolved Sun setup is the fifths iteration with all previous version having different goals and objectives. This latest design was developed with the following goals and constraints in mind:
\begin{itemize}
    \item Re-image the solar disc onto the $525\,\mu m$\,fiber. 
    \item Use a camera setup to image the solar disc and the fiber pickup in order to guide the telescope.
    \item Ideally this should also work on the Moon and on stars.
    \item The design needs to be compact enough to use the limited installation space between the vacuum tube and the wall behind it.
    \item If possible use existing hardware to minimize additional costs.
\end{itemize}

\begin{figure}[H]
  \centering
  \fbox{\includegraphics[width=1\linewidth]{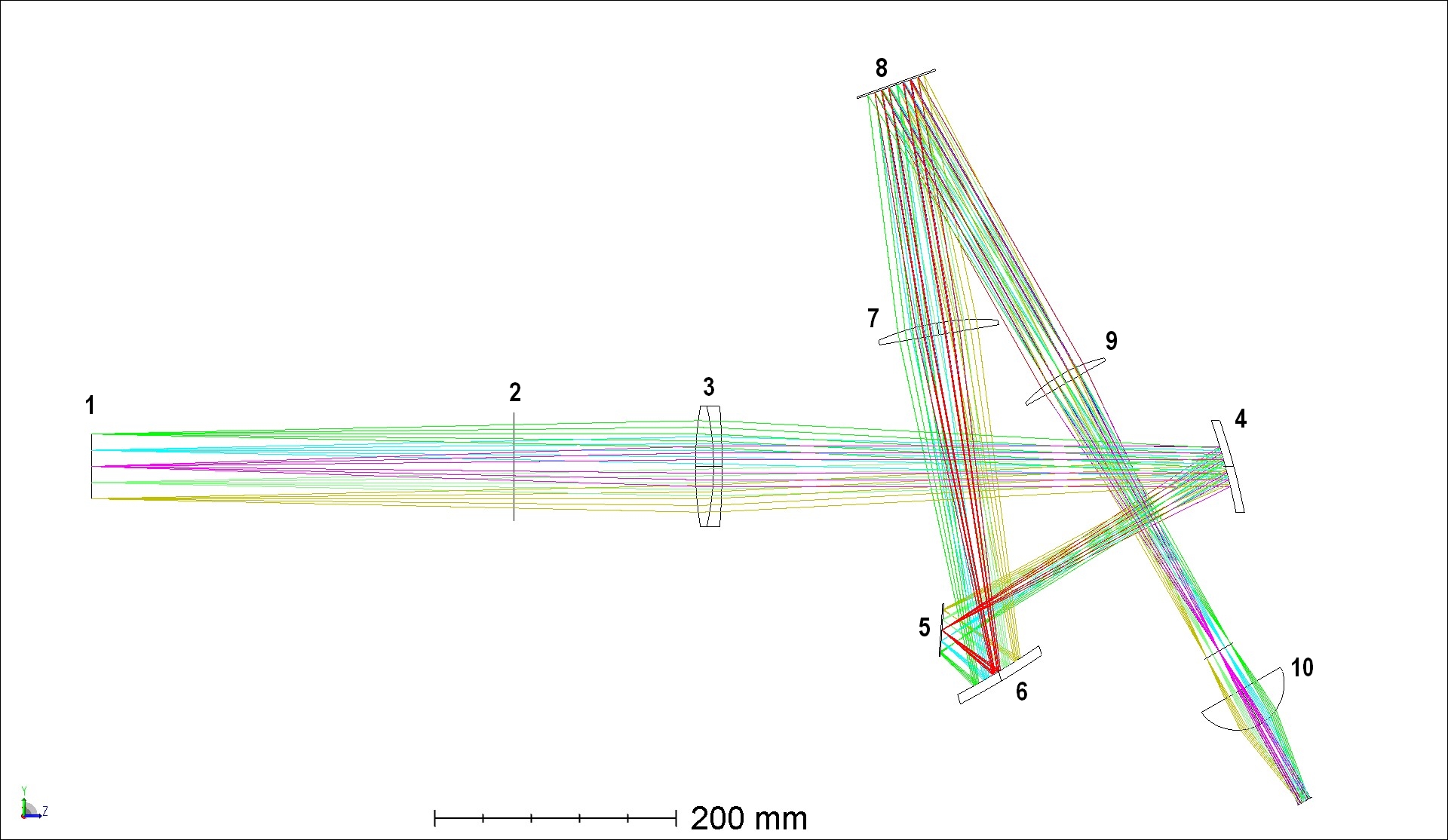}}
  \caption{Zemax simulation of the resolved Sun setup, starting at the Nasmyth focus (1) and ending at the camera position (not shown) behind the last lens (10).}
 \label{fig:zemax}%
\end{figure}

Figure\,\ref{fig:zemax}\, illustrates the optical setup: After the Nasmyth focus (1) the light passes the vacuum window (2). An achromat (3) (Linos G322316000) collimates the light  and a 30\,\degree off-axis parabolic mirror (4) (d=76.2\,mm, EFL=272.2\,mm, Edmund Optics 83963) re-images the solar disc onto the fiber pickup (5). The fiber pickup is a circular, polished stainless steel disc with a 2.5\,mm hole in the center. A FC/PC fiber connector is built into this hole from the backside, so the fiber head is sitting 2\,mm behind the surface. The fiber pickup is tilted by 34\,\degree while the hole is tilted by -34\,\degree so the fiber head is parallel to the image plane. 

The fiber leading towards the FTS is hexagonal shaped with a diameter of $525\,\mu m$ (Ceramoptec HEX WF 525). The angular diameter of the Sun is approximately \ang{;;1920}. With the existing optics this results in a re-imaged solar disc on the fiber pickup with a diameter of 31\,mm, therefore the field of view of the fiber is about \ang{;;32.5}.

For subsequent elements after the fiber pickup both the image quality and the spectral information are not of a big concern anymore, because the scientific data is collected by the fiber and analyzed with the FTS while the guiding system is not limited by image quality. Therefore for the rest of the setup towards the camera, using cheaper components is prioritized over image quality, hence we use simple lenses and accept both spherical and chromatic aberrations. Particularly the chromatic effects can be ignored because a narrow band filter is used in front of the detector to avoid oversaturation. 

After the fiber pickup another off-axis parabolic mirror (6) (d=101.6\,mm, EFL=304.8\,mm, Edmund Optics 83967) and a lens (7) (d=100\,mm, f=402\,mm, unknown manufacturer) collimate the beam. Because the siderostat controls are used in order to move the solar image on the fiber pickup with respect to the fiber, the off-axis parabolic mirror placed behind the fiber pickup needs to be quite close to the fiber pickup. Otherwise the light would be wandering off the mirror when a position on the solar edge is placed onto the fiber. 

A flat mirror (8) folds the light back and a combination of two lenses (9+10) (d=75\,mm, f=250\,mm and d=75\,mm, f=75\,mm, Edmund Optics 86914+86911) is used to compress the beam to a diameter of d=14\,mm. A H-alpha filter and a reflective neutral density filter (not shown) are used to remove most of the light. They can be moved out of the beam for observation of the moon or stars. Finally, a wide angle low distortion lens (not shown) (Edmund Optics 68671) is used in combination with the CMOS detector (The Imaging Source DMK 23GP031). 

\begin{figure}[H]
  \centering
  \fbox{\includegraphics[width=0.5\linewidth]{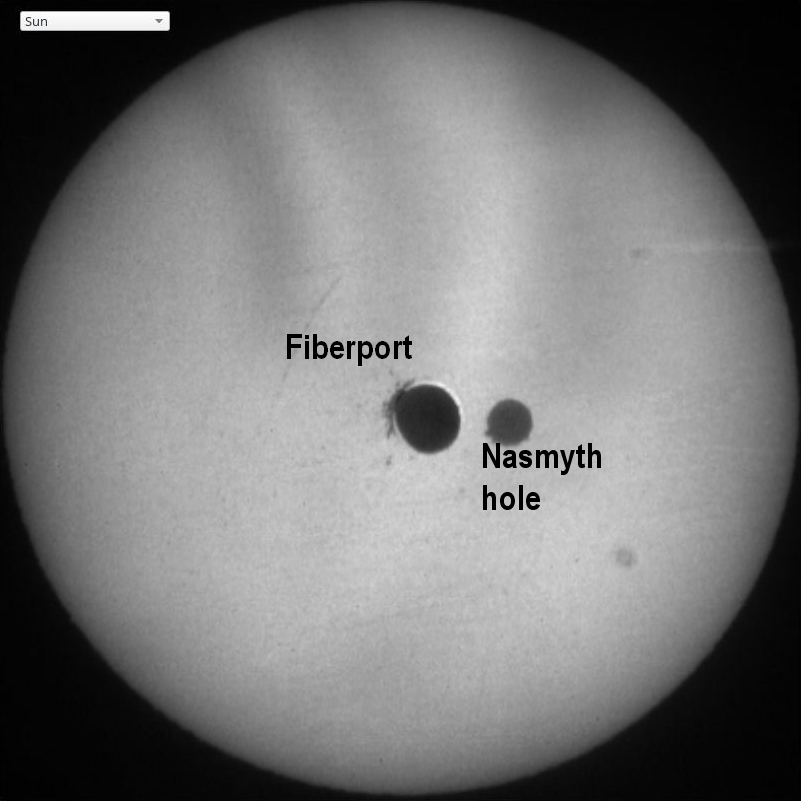}}
  \caption{The solar disc as seen by the detector (including flat-field correction). Both, the hole from the Nasmyth mirror and the fiber hole are visible as shadows on the Sun. The fiber itself is about 1/5 the size of the fiber hole. The remaining shadows and speckles are scratches on the fibre-pickup mirror (5) itself that couldn't be removed completely by the flat-field correction.}
 \label{fig:Sun1}%
\end{figure}

Figure\,\ref{fig:Sun1}\, shows the solar disc as seen by the detector. Measuring the position of the fiber hole, the position of the disc center and a given set of solar coordinates the telescope can be moved in order to hold the region of interest inside the fiber (see Sec.\,\ref{Sec:Software}).

\subsection{First results}
\label{Sec:resolved_results}
\begin{figure}[H]
  \centering
  \fbox{\includegraphics[width=0.6\linewidth]{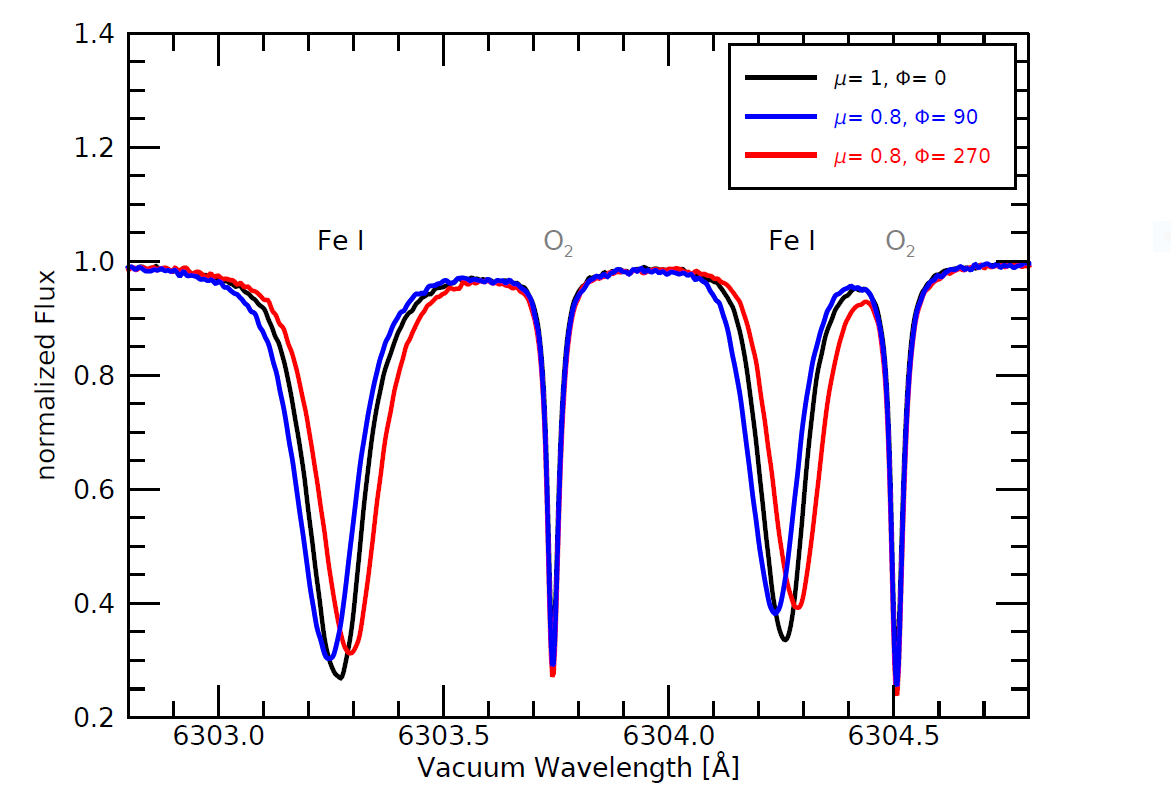}}
  \caption{Segment of the solar spectrum plotted in vacuum wavelength. Spectra for three different areas of the solar disc are shown and the two Fe I solar lines and two O$_2$ telluric lines are marked. The red- and blueshift due to the solar rotation can be clearly seen in the spectra taken towards the solar edge: While the Fe I lines are shifted the telluric O2 lines are not. }
 \label{fig:spec}%
\end{figure}

\begin{figure}[H]
\begin{minipage}{0.48\textwidth}
    \begin{boxedminipage}[t][7.7cm][c]{8.5cm}
		\includegraphics[width=\textwidth]{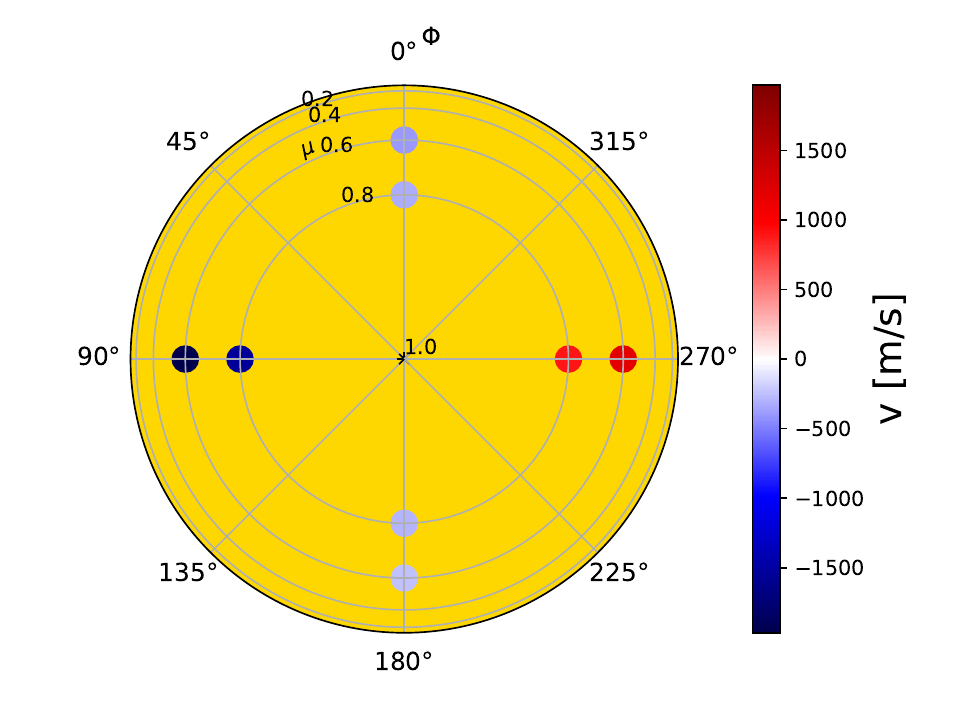}
    \end{boxedminipage}
    \caption{Radial velocities of eight solar surface regions referring to the solar center.}
	\label{fig:rotation}
\end{minipage}
	%\hspace{-8pt}
	~~
\begin{minipage}{0.48\textwidth}
    \vspace{12pt}
    \begin{boxedminipage}[t][7.7cm][c]{8.5cm}
		\includegraphics[width=\textwidth]{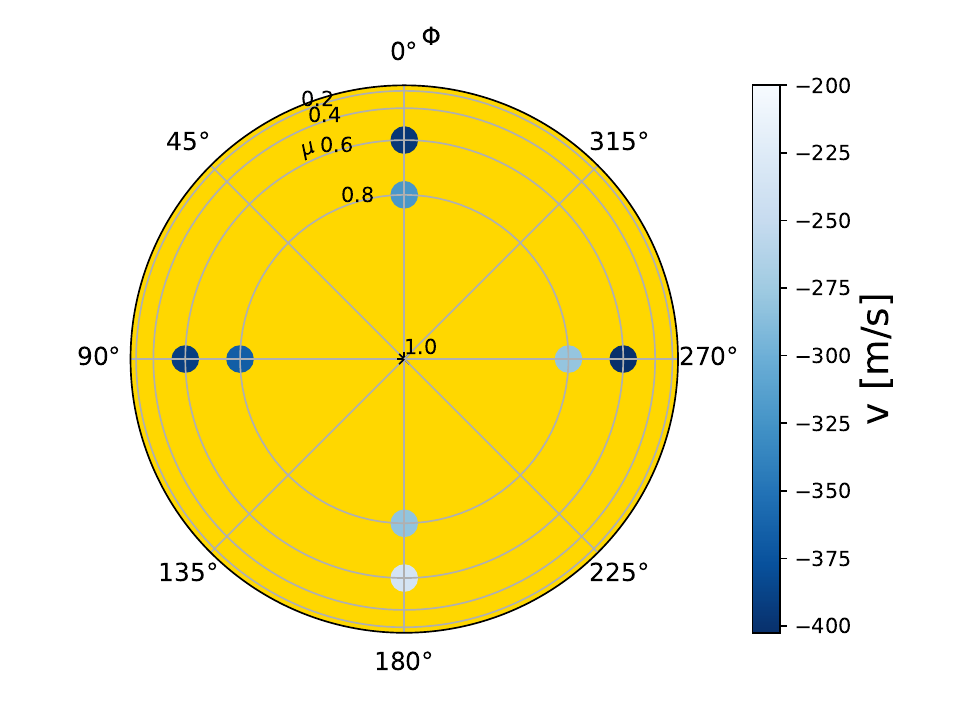}
	\end{boxedminipage}
	\caption{Remaining convective blueshift velocities of eight solar surface regions referring to the solar center after removing the solar rotation.}
	\label{fig:blueshift}
\end{minipage}
\end{figure}

Since the Sun is the only Star we can resolve spatially, it is a matter of particular interest to create a detailed profile of it. For first analysis we use spectra from 9 different areas on the Sun. This data-set was observed on the 8$^{th}$ of April 2020. In Fig. \ref{fig:spec} small segments of three different spectra are shown. This region includes two solar Fe I lines and two telluric O$_2$ lines which are generated by earths atmosphere. The black spectrum corresponds to the solar center ($\mu = 1.0, \Phi = 0$), while the blue one is taken in the east (at $\mu = 0.8, \Phi = 90$) and shows a blueshift of the solar lines based on the solar rotation. On the other hand the red spectrum which is observed in the west ($\mu = 0.8, \Phi = 270$) of the solar surface shows a redshift (cf. Fig. \ref{fig:rotation}). The telluric O$_2$ lines are stable since the motion on the Sun is not affecting them.

 In the data reduction the spectra are normalized and all telluric lines are removed by masking. After this we perform a  cross correlation for the wavelength range of \SIrange{505}{680}{nm}  with the spectra from the center of the Sun used as reference. Afterwards, the barycentric motion between the telescope and the solar region needs to be removed. Therefore we use an ephemeris code \citep{2015PhDT.......200D} which is based on NASA's Navigation and Ancillary Information Facility SPICE toolkit \citep{1996P&SS...44...65A} with an accuracy of a few cm/s. The resulting velocities are shown in Fig.\,\ref{fig:rotation}. As to be expected, the radial velocities are dominated by the solar rotation. Since the solar center serves as reference there is no velocity marker set for this area.
 
We substract the solar rotation using the solar rotation model from \cite{1990ApJ...351..309S} and obtain the convective blueshift velocities which are shown in Fig.\,\ref{fig:blueshift}. These velocities have values between -200 m/s and -400 m/s and are consistent with e.g. \cite{2019A&A...624A..57L}, who calculate the blueshift for different line depths. In their work, they determine the blueshift for 26 different photospheric to chromospheric spectral lines for the four solar axis. In our case the cross correlation relates to roughly 1000 spectral lines of different atmospheric heights. Therefore it is reasonable that our velocities are in the range of the blueshifts of lines between 0.3 and 0.6 in the normalized spectrum of \cite{2019A&A...624A..57L} since we basically measure the mean velocity.

\section{Guiding resolved sun features}
\label{Sec:Software}

The goal of the resolved Sun guiding system is to guide a certain feature on the Sun's surface towards the spectrograph fiber (see Fig.\,\ref{fig:Sun1}) for the duration of the measurement with a $2\sigma$ precision of the fiber-size: \ang{;;32.5}. Since the tracking of the telescope's  original software is not accurate and stable enough to accomplish that, we present an image-based guiding, specifically designed to reduce errors for a spatially extended target (the Sun). A control loop minimizes the distance of the target and the fiber by correcting the position of the S1 mirror (see Fig.\,\ref{fig:Roof}) during  observation. For that, the position of any sun feature in pixel coordinates through coordinate transformations is computed in section \ref{Sec:coordtrans}. Section \ref{Sec:errorcalc} presents a model for the pointing uncertainty.

\subsection{Guiding}
The guiding's control loop tries to minimize the distance between the target position in the image - the \emph{process variable} - and the position of the fiber in the image - the \emph{set point}. The image like seen in Fig.\,\ref{fig:Sun1} provides several important information for image-based guiding.
\begin{itemize}
	\item The \textbf{position of the center of the solar disk} provides a reference point with known position in image and sky coordinates.
	\item The \textbf{ellipsoidal shape of the solar disk} in the image linked with its shape in the  sky provides the scaling between both coordinate systems.
	\item The \textbf{shadow of the fiber hole} provides an estimate for the \emph{set point} of the guiding control loop. %fiber = br, fiber = american, in diesem Fall bitte am benutzen
\end{itemize}
Two important features are however missing from the image.
\begin{enumerate}
	\item The \textbf{exact position of the fiber} relative to its shadow in the image is not known. The fiber is five times smaller than the shadow and might very well not lie at the centre of the shadow.
	\item The \textbf{orientation of the solar disk} in the image is not detectable because it lacks any features visible in the image. Solar spots are too rare during the inactive phase of the solar cycle.
\end{enumerate}
The missing fiber position results in an inaccuracy in the \emph{set point} for the control loop of the guiding process. The missing orientation means that the angle of the target relative to the sun center is to be calculated from the theoretical light propagation as described in Sec.\,\ref{Sec:img2sky}.

The information available in the image is used to build a control loop consisting of the following steps.
\begin{enumerate}
\item
  Detect position and shape of the solar disk in the image (see Sec.\,\ref{Sec:sundetection}).
\item Use the latter to refine the current mapping between Sun and image coordinates (see Sec.\,\ref{Sec:coordtrans}).
\item
  With the updated mapping transform the offset between the center of sun and the target feature at the suns surface to image coordinates.
\item Calculate the \emph{process variable}: the absolute position of the target in image coordinates. This is given by the sum of the detected position of the suns center and the offset.
\item Calculate the difference between fiber and target position.
\item Transform the difference to sky coordinates using the inverse mapping from step two to correct the telescope position.
\end{enumerate}

Steps 1, 2 and 5 introduce uncertainties originating from the sun detection algorithm, the calibration of the optical path and the position of the fibre, respectively. The overall effect of these uncertainties on the guiding result is estimated in Sec.\,\ref{sec:overallaccuracy}.

\subsection{Frame transformation: Sun to Image} \label{Sec:coordtrans}
\begin{figure}
  \centering
  \includegraphics[width=0.8
  \linewidth]{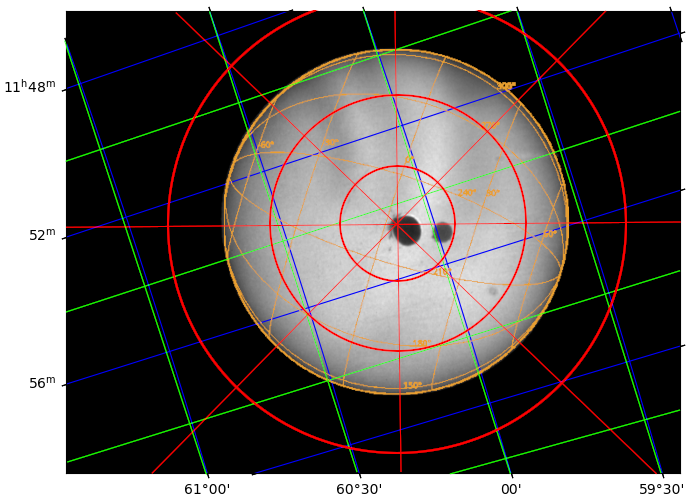}
  \caption{Schematic view of the different coordinate systems used during frame transformations. Orange: Sun's (lon, lat)-coordinates. Blue (curved): absolute topocentric coordinates. Green: Projected Plane coordinates. Red: Native coordinates.}
 \label{fig:overlaysun}%
\end{figure}

Computing the camera pixel coordinates of a sun-feature is the result of a mapping chain that transforms some solar-surface input coordinates. That chain is subdivided into two main transformations. Astrometric mappings transform solar surface (lon, lat)-coordinates to local sky or topocentric coordinates (az, alt) and are realized through implementation of Astropy \citep{astropy:2018}. The instrumental mappings, which concern the transformation of local sky coordinates throughout the optical elements into the guiding-camera, are modelled and fitted to a series of observations. This whole process can be visualized as "drawing" sun's (lon, lat)-coordinates within the sun's disk onto the camera image like shown in Fig.\,\ref{fig:overlaysun}. Appying these transformations introduces some uncertainties which are investigated closer in Sec.\,\ref{Sec:errorcalc}.

We first introduce the following coordinate systems, and refer to their definition for further reading.
\subsubsection{Reference frame definitions}
\label{sssec:framedef}

\begin{itemize}
\item \textbf{Helioprojective radial (HPR) coordinates ($\mu$, $\phi$)}
Commonly used in solar science, it defines a two dimensional polar frame on the visible sun's disk, from the observers perspective \citep{2007CeMDA..98..155S}. It is, together with Stonyhurst (see below), one of the solar frames in which user input occurs.

\item \textbf{Stonyhurst (SH) coordinates (lon, lat)}
In SH coordinates, the longitude starts counting from the central meridian \citep{2007CeMDA..98..155S}, which is the plane spanned by the observer, the sun center and the sun's north pole. The latitude starts counting positively towards the sun's north pole.

\item \textbf{Refracted topocentric / Sky coordinates (az, alt)}
 Describe the position, corrected by atmospheric refraction, of a target within the local sky above the telescope. They constitute the "bridge frame" between astrometric and instrumental mappings. We follow the World Coordinate System and FITS convention \citep{2002A&A...395.1077C} to transform celestial coordinates to pixel coordinates. In the latter section, we refer to them as celestial \textbf{"Real World Coordinates (RWC)"}.

\item \textbf{True camera coordinates (x, y)}
As the last frame in the chain, it provides the guiding system with empirically measured properties of the sun like the position and shape of the solar disk in the camera image.

\item \textbf{ICRS coordinates(ra, dec)}
Solar system barycentric, equatorial aligned coordinate system  \citep{1998AJ....116..516M}. HCRS and GCRS correspond to the translation of the system to the heliocenter and geocenter, respectively.

\item\textbf{Celestial Intermediate Reference System (CIRS) (ra, dec)}
CIRS are the modern equatorial "of Date" coordinates, using the Celestial Intermediate Origin (CIO) instead of the equinox of date. Catalogue coordinates are generally corrected by proper motion and parallax, gravitational deflection of light, precession and nutation  \citep{2008IAUS..248..367C}.
\end{itemize}

\subsubsection{Astrometric mapping: Sun to sky}
\label{Sec:img2sky}
\begin{figure}[h]
  \centering
  \includegraphics[width=1
  \linewidth]{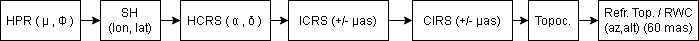}
  
  \caption{
  Astrometric coordinate transformation steps from solar Helioprojective radial to Refracted topocentric coordinates.}
 \label{fig:astrom_subchain}
\end{figure}

We aim to compile a coordinate transformation chain that returns refracted topocentric $(az, alt)$, telescope centered coordinates of a target on the Sun's surface at a certain time. Through the Python library Astropy, we can use software routines from The Standards of Fundamental Astronomy (\cite{SOFA:2020-07-21}), which is maintained by the International Astronomical Union. They provide all  ephemeris methods and astrometric transformations that are needed for such a task. Sunpy \citep{sunpycommunity2020} is used for solar coordinates. The uncertainty propagated by this chain will be neglected, since its errors ($\pm \ang{;;0.06}$) lie some orders of magnitude below the instrumental errors. This subchain's structure is shown in Fig.\,\ref{fig:astrom_subchain}.
\hspace{1cm}

\subsubsection{Instrumental mapping: Sky to Pixel} \label{sssec:instrumental}
The goal of the instrumental mapping is to transform topocentric coordinates (relative to the Sun-center) through the optical elements into pixel coordinates. Given (from the astrometric mapping) the spatially extended target (Sun) in the sky as a set of topocentric coordinates and its corresponding Sun-surface coordinates, this actually completes the transformation of the solar surface to pixel coordinates. We use the empirically detected Sun's pixel position and size (see Sec.\,\ref{Sec:sundetection}) in the transformation, adjusting the calculated Sun grid to the observed sun, which saves some modelling effort, but needs the Sun to be in the guiding image.

\begin{figure}[h]
  \centering
  \includegraphics[width=0.8
  \linewidth]{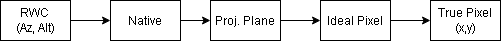}
  \caption{Instrumental coordinate transformation steps from real world coordinates to pixel coordinates in the camera image.}
 \label{fig:intrumental-chain}%
\end{figure}

\begin{figure}[h]
  \centering
  \includegraphics[width=0.9
  \linewidth]{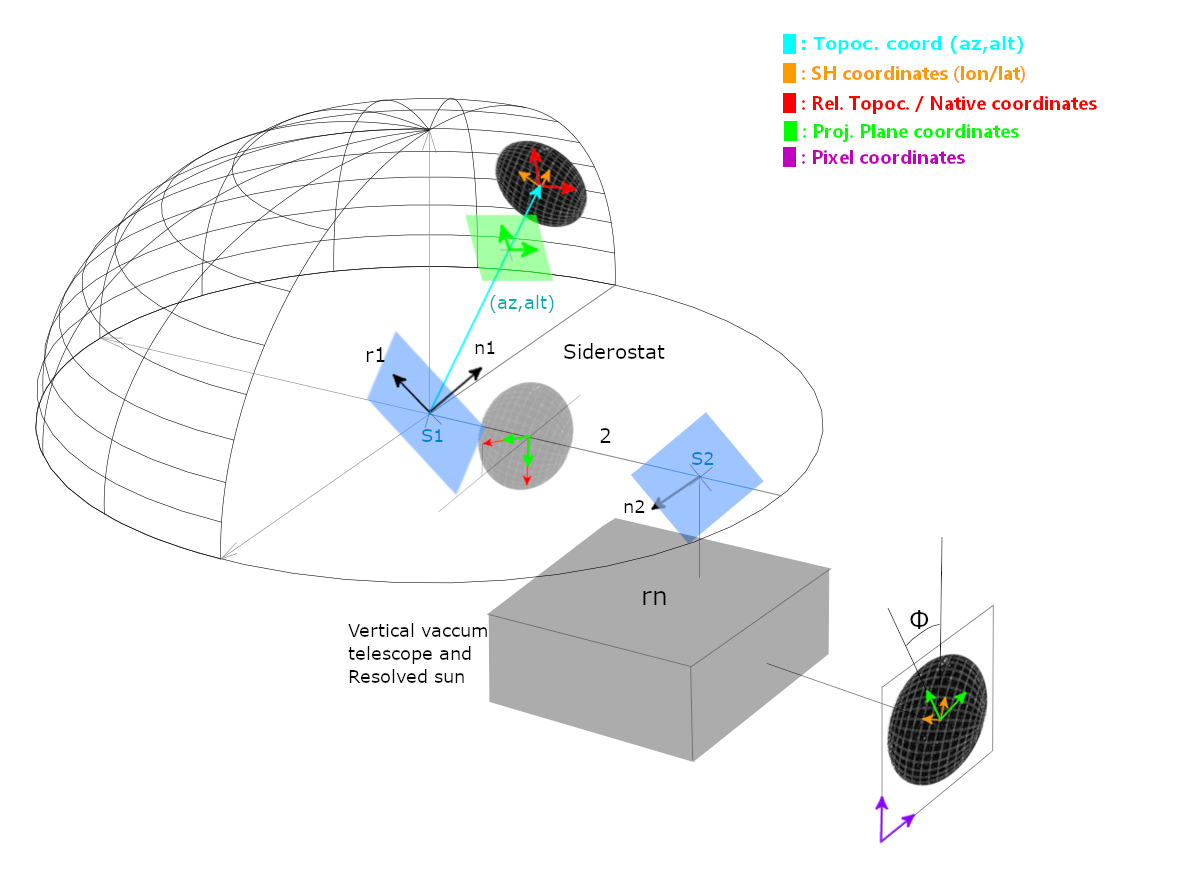}
  \caption{Overview of the siderostat-specific coordinate transformations for a spatially extended target (Sun or Moon). See Figs.\,\ref{fig:overlaysun} and \ref{fig:intrumental-chain} for comparison.}
 \label{fig:half-sky}%
\end{figure}

This subchain follows World Coordinate Systems (WCS) transformation conventions \citep{2002A&A...395.1077C}: For a ground-based instrument, we refer to topocentric/sky coordinates as celestial "Real World Coordinates (RWC)" in WCS terminology. Figure\,\ref{fig:intrumental-chain}\, gives a schematic overview over all the instrumental transformation steps and Fig.\,\ref{fig:half-sky} illustrates them in a 3D context.

Due to the small telescope's field of view (FOV) ($D \approx \ang{0.5}$), this subchain has been approximated by a linear map in the past, creating a systematic error.

\begin{enumerate}
    \item \textbf{Real World (az, alt) to Native coordinates}
This process corresponds to rotating the pole of spherical topocentric coordinates towards the Sun-center, which serves as a reference point for all the following transformations  \citep{2002A&A...395.1077C}. When computing these (Sun-center relative) coordinates, we cancel any error concerning absolute Sun-center's position (especially atmospheric refraction and uncertainty about telescope's location\footnote{A (lon, lat) error would impact 1:1 the absolute topocentric coordinate error}), leaving us to deal only with errors concerning the orientation of the Sun's sphere. This can be done, since we obtain the Sun-center in pixel empirically by fitting an ellipse in the camera image rather than tracking its absolute position in the sky. However, a new error detecting the Sun in pixel is being introduced here, which we will discuss later.

\item \textbf{Native to Projection Plane coordinates}

Next step is to project the native sphere into the (cartesian) projection plane (in units of arcsec; orthogonal to the optical axis for any telescope position). For an observer in the center of the celestial sphere, this is called a gnomonic projection. For this step we use premade  methods within astropy.

\item \textbf{Projection Plane coordinates to Ideal Pixel coordinates:}

The transformation from projection plane coordinates throughout the whole optical system and into the camera pixel coordinates. This is, in the case of a perfectly aligned optical system, a linear transformation \(T=R\lambda\) composed of a $2\times 2$ dimensional rotation/roll matrix $R$ and a physical scaling from arcsec to pixel represented by a diagonal matrix $\lambda$. We will discuss these two separately, as we can again use the fact that our target is spatially extended to avoid inserting new pointing uncertainties.

\begin{itemize}
\item \textbf{Scaling arcsec to ideal pixel (circular sun):} \label{scal1}Instead of modelling the different focal lengths and the distances of the optical elements to obtain the scaling factor (which can deviate from the physical one), we empirically measure the circular pixel size of the Sun within the guiding camera. We also know, thanks to Astropy's ephemerides, its angular size at any given time. The ratio of these two provides us with the scaling factor and diagonal matrix \(\lambda\) we are looking for. But this is only half the truth for scaling, as it will see in the last transformation to true pixel coordinates.

\item \textbf{Modelling the optical path's reflections to obtain the roll angle:}
A first approach is to simply model every optical element's reflection by applying a three dimensional rotation in each step. We orient both axis of the Projection Plane in three dimensional space (x-axis must be parallel to the ground), and rotate them around the axis \(r_1\) (normal to this step's reflection plane) that describes the orientation of the first optical element, the S1 mirror in this case (see Fig.\,\ref{fig:half-sky}). This provides us with the orientation of the reflected Proj. Plane system (which already contains the Sun-surface system) in optical path 2 between S1 and S2 . This is repeated for every optical element, always using nominal (theoretical) values \(r_n\) for every mirror's orientation. At last we project the result into the two dimensional frame of the camera and read the modelled orientation \(\phi\) of the Projection Plane system in the camera image. This angle only depends on the S1 mirror's orientation (which is unique for a certain Sun position \((az,alt)\), therefore we can define a model function that maps topocentric coordinates to a roll angle \(f:(az,alt)\to\phi\).

\item \textbf{Resolved Sun setup orientation calibration - Fitting the model to observations of the sun, bright stars and landmarks:}
The true orientation of the optical elements does not exactly match the nominal values, which causes a deviation between measured roll \(\phi_n\) at \((az,alt)_n\) and modelled roll angle \(f(az,alt)=\phi\). Therefore, we use the sun, some bright stars and landmarks on the horizon to actually measure the roll angle. For that, we make small movements \((\delta az,\delta alt)\) and measure the angle \(\phi_n\) of the resulting image shift in pixel coordinates \((\delta x,\delta y)\). We obtain a dataset with different sky positions and according roll angles \(\{az_n, alt_n, \phi_n\}\). Then we fit our model, leaving every orientation \(r_n\) as a free parameter. Figure\,\ref{fig:fitresiduals} shows the residuals of that fit. From the standard deviation of the residuals we obtain the uncertainty of the orientation as listed in Table\,\ref{tab:uncertainty}.
\end{itemize}

\begin{figure}[h]
  \centering
  \includegraphics[width=1\textwidth]{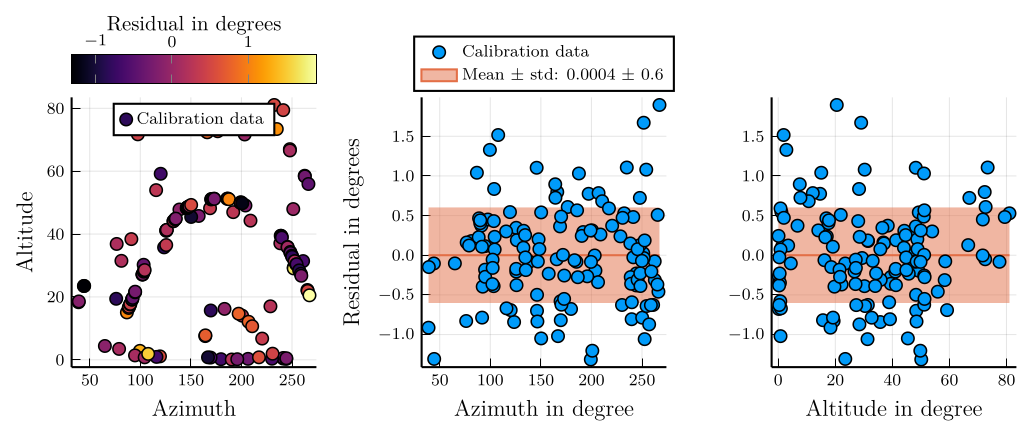}
  \caption{For several positions at the sky, represented by the points in the left image, measurements of the roll angle \(\phi_n\) were made. This allows to fit our \(f:(az,alt)\to\phi\) model, that describes the relative orientation of image and sky coordinates. The residuals of that fit plotted against the local sky coordinates in the middle and right image show no obvious systematics, which can also be seen in the color coded residuals in the left image. Together with a mean value of zero for the residuals (red line in middle and right image) this means that our model provides a good description of the actual orientation. The standard deviation (red area in middle and right plot) provides a measure for the uncertainty in the orientation returned by our model.}
 \label{fig:fitresiduals}%

\end{figure}

\item \textbf{Ideal Pixel coordinates to True Pixel coordinates (ellipsoid)}
At last, we observe an optical distortion of the Sun that makes it appear an ellipsoid rather than the expected circular shape. An extension of the last empirical scaling method is following: First, use some arbitrary circular pixel Sun size for the last scaling ratio (arcsec to ideal pixel, see above). Then run an ellipsoid detection method, which returns its orientation, position and two different axis sizes, and compute the 2x2 transformation matrix relative to that arbitrary ideal pixel size (circular sun).

This means that we fit the whole transformation chain to the empirically observed ellipsoid, sparing a detailed modelling of the instrument and still reaching high pointing accuracy. Nevertheless, the ellipsoid detection itself carries uncertainties, concerning size and position.

Sun-detection accuracy: To get an estimate of the Sun detection's uncertainty we simulate several Sun images. We run the principle component analysis based ellipsoid detection method (see Sec.\,\ref{Sec:sundetection}) on images with different Sun positions, sizes, angles and levels of noise and compare its results to the simulated values. The uncertainty displayed in Table\,\ref{tab:uncertainty} is determined from its stochastic evaluation.

Note: these values can only be considered rough estimates of the algorithm's performance for real images. On the assumption that the actual uncertainty about the Sun-center's position in image coordinates is actually isotropic, we choose the biggest value and call each cartesian component \(\sigma_{pos}\) for the error  analysis below\footnote{One reason to have measured different values could be the fiber's shadow affecting the detetion in a non-symmetric way.}.We apply the same for the detected size and fiber position uncertainty \(\sigma_{size}\) and \(\sigma_{fib}\).
\end{enumerate}

\subsection{Overall pointing uncertainty} \label{Sec:errorcalc}
\label{sec:overallaccuracy}
\begin{table}
    \centering
    \caption{Estimated partial uncertainties collected along the transformation chain.}
    \begin{tabular}{rc}
        \toprule
        Source of error & estimated uncertainty \\
        \midrule
        Detected position of Sun center $\sigma_\text{pos}$ & \ang{;;1.10} \\
        Detected size of Sun $\sigma_\text{size}$ & \ang{;;3.67} \\
        fiber position $\sigma_\text{fib}$&  \ang{;;3.00}\\
        %fiber position $\sigma_\text{fib}$&  \SI{0.005}{Sun radii} \\
        Calibrated orientation of Sun $\sigma_\text{roll}$ & \ang{0.6}\\
        \bottomrule
    \end{tabular}
    \label{tab:uncertainty}
\end{table}

In the following, linear error propagation is applied to the instrumental mappings involved in the guiding procedure. We aim to express the pointing uncertainty as a function of the target's position in helioprojective radial coordinates. All partial uncertainties/errors that contribute in a significant way to the overall uncertainty are displayed in Table \,\ref{tab:uncertainty}. As stated above, we neglect the errors from the astrometric mappings. We then compare it with actual pointing observations on some moon features, which serve as reference targets on a spatially extended body.

Before doing that, we introduce the estimated uncertainty of the fiber's position in the pixel coords. (see Table\,\ref{tab:uncertainty}), which has still to be determined in the future. An isotropic uncertainty is assumed also at this step. Note that, although we use a yearly averaged Sun radii of $r_{\text{\astrosun}} = \ang{;;965}$ for the arcsec - Sun radii conversion, it actually oscillates with an amplitude of \ang{;;16} over the year. The parameter $r$ expresses the unitless distance of the target from the center of the visible Sun's disk in an interval $[0,1]$.

First step is to merge the fiber's position uncertainty and the detected Sun-center's position uncertainty through standard-deviation sum, because they both contribute as an isotropic uncertainty (independent of the target's position on the Sun) to the overall uncertainty. Each of its components reads:

\begin{align}
\sigma_\text{off} &= \sqrt{\sigma_\text{fib}^2+\sigma_\text{pos}^2}
\end{align}
We map the roll angle uncertainty from Table\,\ref{tab:uncertainty} to the corresponding tangential uncertainty as a function of radial distance $r$. The Sun center is the reference point/pole of the linear transformation and is therefore free from this error (see Sec.\,\ref{sssec:instrumental}). The scaling uncertainty causes a radial error, again depending linearly on the radius.

\begin{align}
    \sigma_\text{tan}(r) &= \sin{(\sigma_\text{roll})}\ r \ r_{\text{\astrosun}} \ \label{eqn:tan} \\
    \sigma_\text{rad}(r) &=
    \sigma_{size} \ r \label{eqn:rad} \ 
\end{align}
Merging all uncertainties results in the component-wise errors:
\begin{align}
    \sigma_\text{total, tan}(r) &= \sqrt{\sigma_\text{off}^2 +
    \sigma_\text{tan}(r)^2}
    \\
    \sigma_\text{total, rad}(r) &= \sqrt{\sigma_\text{off}^2 + 
    \sigma_\text{rad}(r)^2}. 
\end{align}

The absolute 1 $\sigma_\text{total}$ distance follows from:
\begin{align}
    \sigma_\text{total}(r) &= \sqrt{\sigma_\text{total, ang}(r)^2 + \sigma_\text{total, rad}(r)^2}.
\end{align}

Any of the error components, as well as the absolute $\sigma_\text{total}$ can be expressed as a function of Helioprojective radial (HPR) coordinates considering that they are $\phi$-independent and applying:

\begin{align}
    \mu=sin(cos^{-1}(r)) = \sqrt{1 - r^2}.
\end{align}

\begin{figure}
  \centering
  \input{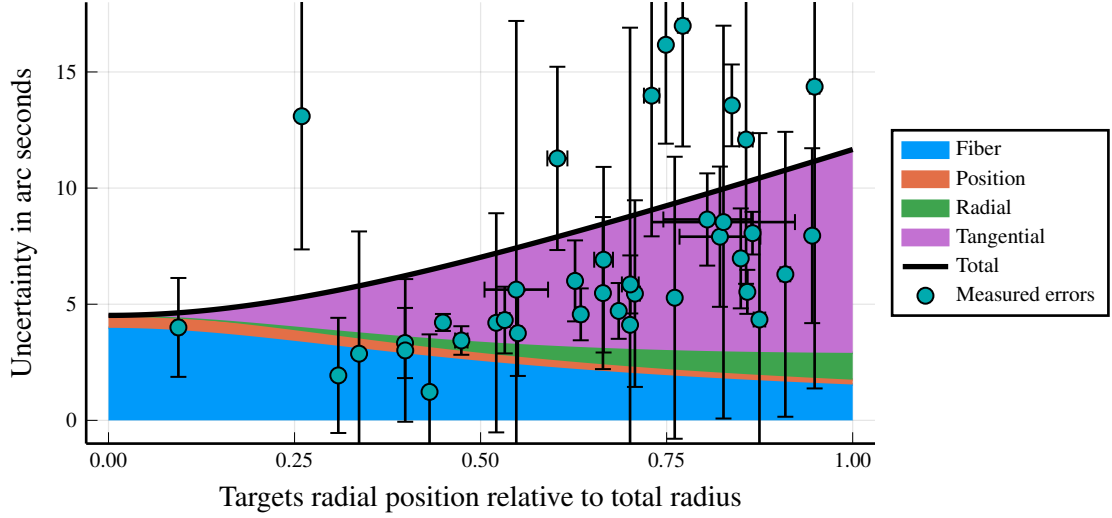}
  \caption{Comparison between measured pointing errors from moon observations (blue points) and analytical uncertainty function from linear error propagation (black line). The latter results from the partial uncertainties. Their contribution to the total uncertainty is displayed as a stacked area plot. Each of the blue markers represents a specific target at the lunar surface. Their mean position and errorbar result from multiple measurements for that target.}
 \label{fig:errorplot}%
\end{figure}

To compare this function with real measurements, we used contrast-rich features on the moon's surface as a reference. A target's coordinates in $(lon, lat)$\footnote{Of course, some astrometric transformations had to be adapted to the moon's selenographic coordinates.} are fed into the guiding system, and several target images are taken. Through comparison/correlation with simulated Moon images by NASA Scientific Visualizations Studio's project "Dial-a-Moon" (see \cite{SMITH201770} and \cite{Sato2014}), we measure the mean and standard deviation of the pointing error for that target. 
Figure\,\ref{fig:errorplot} shows the pointing error results for 38 moon targets against the radial distance $r$, along with its error bars in radial position (horizontal) and in absolute pointing error (vertical). The partial uncertainty's squares divided by the total error are plotted as stacked areas. Their sum (black line) represents the total 1 $\sigma_\text{total}$ distance of the pointing error. Near the Sun-center the fiber position uncertainty dominates the total uncertainty, while the tangential uncertainty grows to be the largest fraction for targets close to the limb. The radial and (Sun-detection) position errors are comparably small over the whole range. These systematics will be useful for further pointing improvements in the future. The measured errors from the Moon observations are in good agreement with the analytic uncertainty function.

Throughout a spatially extended target, the maximal 1 $\sigma_\text{total}$ distance is about \ang{;;12.00}. This means even with a 2 $\sigma_\text{total}$ of \ang{;;24.00} we are still well inside the required \ang{;;32.50} of the fiber diameter.

\newpage
\subsection{Detection of solar disk}
\label{Sec:sundetection}
The aim of the detection algorithm for the solar disk is to provide reliable information about its shape and the position in the image. The ideal image of the solar disk is an evenly illuminated circle against a black background. However, there are several aspects changing that ideal image.
\begin{itemize}
    \item \textbf{Limb darkening} towards the border of the solar disk.
    \item \textbf{Atmospheric refraction} flattens the shape in direction of altitude.
    \item \textbf{Optical distortion} of the ideal circle to a rotated ellipse (or worse).
    \item \textbf{Atmospheric turbulence} causes distortions mainly visible as a wobbling in the contour of the disk.
    \item \textbf{Static shadows} in the image caused by the limited size of the fiber-pickup mirror and the hole for the fiber in it.
    \item \textbf{Static attenuation} in intensity from inhomogeneities at the surface of the fiber-pickup mirror.
    \item \textbf{Dynamic shadows} caused by the hole in the Nasmyth mirror and the spider fixing it inside the vertical vacuum tube and clouds, birds and airplanes in front of the sun.
\end{itemize}

These effect need to be mitigated by several steps:
\begin{itemize}
    \item \textbf{Careful design} of the current optical setup makes sure that shadows from the spider only result in a spatially homogeneous change in the images intensities. We also make sure that always the whole solar disk fits onto the mirror of the fiber-pickup and thus avoiding cut offs from its limited size.
    \item Use of \textbf{flat-field correction} reduces the effect from static attenuation in the image.
    \item \textbf{Masking} the area in the image with the shadow from the hole in the fiber pickup allows ignoring the influence of those pixel for the detection. Since the spatial movement of the shadow of the hole in the Nasmyth mirror is only small the mask can cover that area too.
    \item \textbf{Thresholding} the image intensities separates the solar disk from the background and removes the effect from limb darkening.
    \item \textbf{Averaging} over multiple images reduces the effect from atmospheric turbulence.
\end{itemize}

Ignoring clouds, this leaves only the distortion in shape from atmospheric refraction and distortions in the optical path. For simplicity we assume that both only introduce linear distortions resulting in the shape of a rotated ellipse which we need to detect. Note however that the nonlinearity introduced by the distortions can be very well non-negligible, especially when observing low altitudes.

We implement two different detection methods. The first, based on principal component analyses is very simple to implement and stable in the sense that consecutive detection results never differ by much thus avoiding sudden jumps in the guiding process. The second method based on edge detection and least squares fit of an ellipse is more accurate but can give no or completely wrong results for large deviations in the image, for example from clouds.

Both of the methods will be described in more detail in the following:

\paragraph{The principal components based method} uses statistics of the point cloud of the coordinates of each illuminated pixel in the threshold image to estimate the ellipse parameters.
\begin{itemize}
    \item The mean of the point cloud is the center of the ellipse.
    \item For an unrotated ellipse the standard deviation $\sigma_x$ and $\sigma_y$ of the point cloud is proportional to the major axis $a=4\sigma_x$ and the minor axis $b=4\sigma_y$.\footnote{This can be checked by calculating the standard deviation of an ellipse along the major axis with length $a$ as $\sigma_x = \sqrt{\frac{1}{\pi a^2} \int_0^a\int_0^{2\pi} (r\cos(\phi))^2 r\text{d}r\text{d}\phi}$ and for the minor axis accordingly.}
    \item Principal component analysis gives the direction with largest spread of the point cloud which can be assigned to the orientation of the ellipse.
\end{itemize}
Based on this, the implemented ellipse calculation contains the following steps\footnote{A similar method, using only pixels on the edge of the ellipse, is described in \cite{pcaellipse}.}.
\begin{enumerate}
    \item Calculate center of the ellipse as the mean of the point cloud.
    \item Calculate the covariance matrix of the mean shifted point cloud.
    \item The eigenvalues and eigenvectors of the covariance matrix respectively give the lengths and directions of the ellipses axes. Note that the eigenvalues give the standard deviation of the point cloud in the direction of the according eigenvector and therefore scale with a factor of four to the length of the ellipses axis.
\end{enumerate}
Since all pixels of the solar disk are included in the statistics, the method is rather robust against small deviations from an ideal rotated ellipse. On the other hand every deviation from the ideal case will introduce an error in the detection result. It is also not clear how masking of the pixels around the shadows of the fiber and the Nasmyth hole could be handled with this method.

\paragraph{The edge detection and direct least squares based method} first extracts edges from the threshold image using Canny edge detection \citep{canny1986computational} implemented in the OpenCV library \citep{opencv_library}. All edge pixel falling into masked areas are removed. To the remaining points an ellipse is fitted using the method described in \cite{ellipsedetection}. While mathematical justification of the method is not trivial its implementation is easy enough, needing only a few lines of code with help of modern linear algebra libraries like NumPy \citep{harris2020array}.

The obvious advantage of the method is that masking is easily applied and defects only affect the detection result if they lie at the contour of the sun. The disadvantages are that getting the parameters of edge detection correct can be difficult and wrong edges that could be caused by clouds can result in completely wrong detection results. This can make the detection much more jumpy than the result from the statistical approach.

\section{Integrated Sun}
\label{Sec:integrated}
\subsection{Former setup}
In 2014 a first attempt to observe the integrated sun using the siderostat and the FTS has been made. In this setup a 12.5\,mm diameter off axis parabolic mirror was placed directly below the S2 mirror to form an image of the full solar disc onto a $800\,\mu m$ fiber which lead to the FTS. This work resulted in the IAG solar flux atlas \citep{Reiners_2016} which showed similar fidelity to the Kitt Peak FTS atalses (\cite{1984sfat.book.....K} and \cite{2011ApJS..195....6W}) and consistent line depths compared to both the Kit Peak and the HARPS \citep{2013Molaro} solar atlases.

The experimental setup was mainly limited by errors originating from tracking errors. Since we image the full solar disc onto the fiber at roughly 2/3 of the fiber diameter the coupling efficiency is changing when the disc moves with respect to the fiber. This mainly concerns the limb areas of the solar disk and we found that this effect can have an amplitude of several $10\,m\,s^{-1}$. In order to improve this the new setup does not image the solar disc onto the fiber head directly but uses an integration sphere to ensure a stable coupling into the fiber. 
\subsection{New setup}
The new setup consists a flat mirror (1) (d=50.4\,mm, protected silver coating) at 45\degree \,that is placed directly below the S2 mirror of the siderostat. Therefore we can still use the siderostat to track the sun while the effective mirror size is reduced from a circular 500\,mm to an elliptical mirror with the minor axis of $d=\sqrt{50.4}$\,mm. The light is send onto an off axis parabolic mirror (2) (d=50.8\,mm, EFL=101.6\,mm, Thorlabs MPD249-P01). This mirror focuses the light into an integrating sphere (3) (d=38.1\,mm, spectralon coating, entrance hole d=9.5\,mm, Ocean Insight FOIS-1) with a fiber exit port where the fiber ($d=800\,\mu m$, NA=0.39) leading to the FTS is plugged in.

\begin{figure}[H]
  \centering
  \fbox{\includegraphics[width=0.9\linewidth]{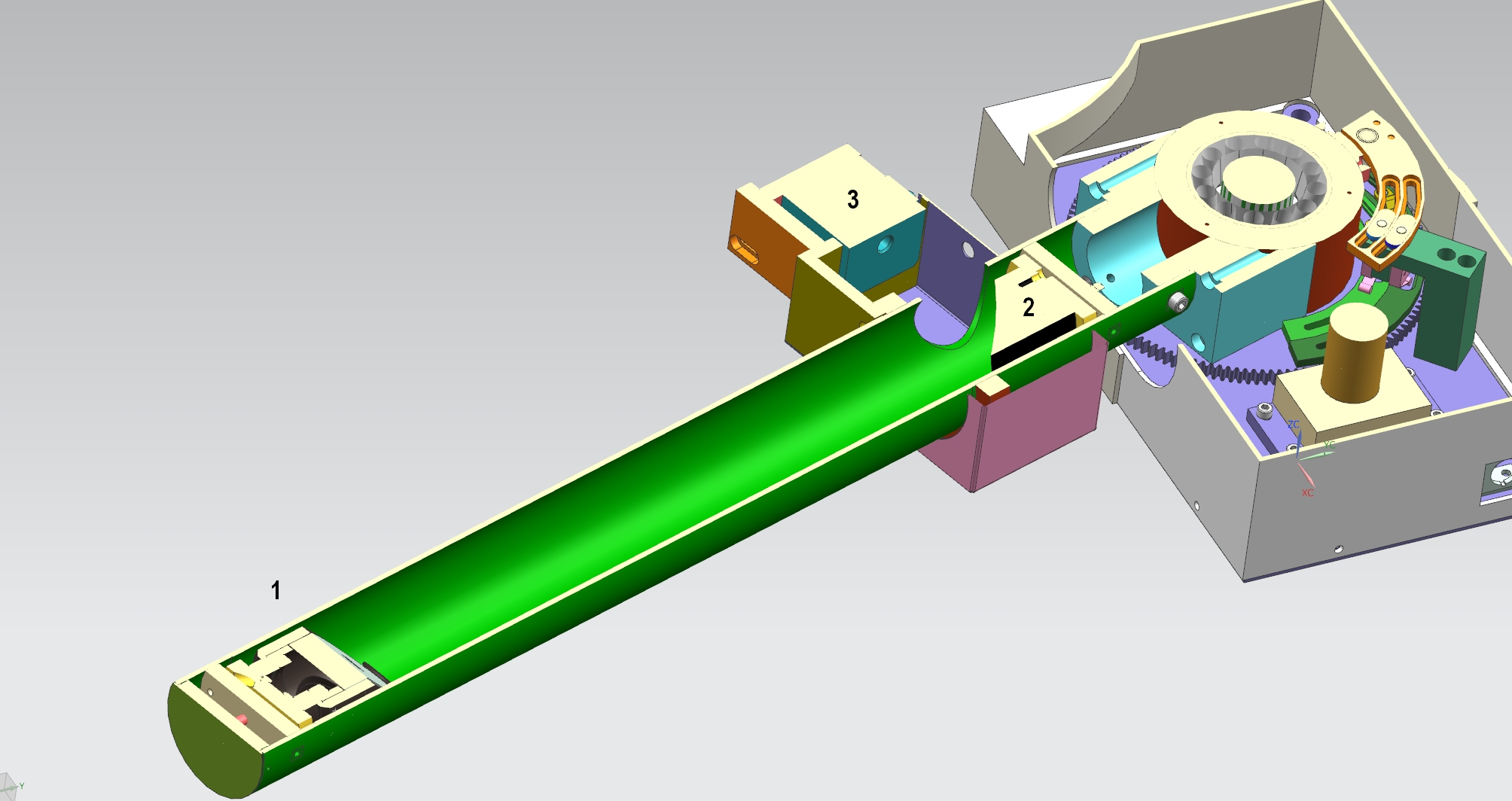}}
  \caption{Movable arm containing the integrated Sun setup: The flat mirror (1) is placed in the center of the beam coming from the S1 and S2 mirrors of the siderostat. Because no focussing optics are in front of it, the light still contains the full disc information. An off-axis parabolic mirror (2) is used to focus the light through the entrance hole of the integrating sphere. The sphere (3) has a fiber exit port (not shown) where the fiber leading to the FTS is plugged in.}
 \label{fig:intergrated_mech}%
\end{figure}

The whole setup is placed inside an aluminium pipe that can be moved by an electric motor in or out of the beam. This means the telescope guiding can still be used to track the sun while the 45\,\degree mirror is always in the center of the beam. Figure\,\ref{fig:intergrated_mech} shows a CAD image of the setup.

\subsection{First results}
The usage of an integrating sphere leads to a significant loss of light due to the many internal reflection, even with special coatings light spectralon. Rough estimates with the given geometry of the sphere results in an overall efficiency of the sphere coupling of \num{4e-4}. At the same time the effective mirror area increased from 491\,mm$^2$ to 5733\,mm$^2$ compared to the old setup. Additionally we also increased the performance of the coupling into the FTS by a factor of 5 (see other publication). In total we lose a factor of about 40 in terms of efficiency. At the same time we also expect to be much less effected by tracking errors because the integrating sphere essentially mixes all spatial features of the solar disc before coupling it into the fiber. This will be investigated in the near future.

\newpage
\section*{Ackowledgments}

% sunpy software
This research used version 2.0.1 \citep{stuart_j_mumford_2020_3940415} of the SunPy open source software package \citep{sunpycommunity2020}, the Astropy library \citep{astropy:2013} \citep{astropy:2018} and NASA's Scientific Visualization Studio.

\bibliography{SPIE2020} % bibliography data in report.bib
\bibliographystyle{plainnat} % makes bibtex use spiebib.bst
\end{document}